**Scott McClellan, MRC/CCI, Drexel University, <u>sm4522@drexel.edu</u>**
**Mat Kelly, MRC/CCI, Drexel University, <u>mrk335@drexel.edu</u>**
**Jane Greenberg, MRC/CCI, Drexel University <u>jg3243@drexel.edu</u>**

# Modeling Ephraim Chambers' Knowledge Structure from a Naïve Standpoint

## Abstract

In the preface to his *Cyclopaedia* published in 1728 Ephraim Chambers offers readers a systematized structure of his attempt to produce a universal repository of human knowledge. Divided into an interconnected taxonomic tree and domain vocabulary, this structure forms the basis of one effort from the Metadata Research Center to study historical ontologies. The knowledge structure is being encoded into a Simple Knowledge Organization System (SKOS) form as well as a Web Ontology Language (OWL) version. This paper explores the expressive and functional differences between these SKOS and OWL versions of Chambers' knowledge structure. As part of this goal, the paper research focused on the construction and application of rules in each system to produce a more computationally ready representation of Chambers' structure in SKOS, which is more thesaurus-like, and OWL, which represents additional ontological nuances. First, studying the various textual aspects at the semantic, syntactic, and typographic levels allowed for the relationships between terms to manifest from which rules governing expression of the connections between elements developed. Second, because each language, SKOS and OWL, functionally expresses different logical relationships, their possibilities and limitations offer a ground for further analyzing the resultant knowledge structures; although, each stemmed from the same basic source of Chambers' text. Lastly this paper will examine rule making and expression in light of Paul Grice's theory of conversational implicature to understand how a naïve agent formulates and applies these rules to a knowledge structure.

## Introduction

Historical knowledge structures give insight into perceptions, beliefs, and knowledge constructs of the past. One example is Ephraim Chambers' *Cyclopaedia,* which had several editions published between 1728 and 1752. The preface of the *Cyclopaedia* includes a bipartite ontology, consisting of a taxonomic tree connected to a domain vocabulary. In an effort to prepare this system for computation, a group of researchers at the Metadata Research Center (MRC) sought to first convert this text into the W3C's Simple Knowledge Organization System (SKOS), a machine-readable encoding language for lexical-semantic thesauri. Initial testing is reported in Greenberg, et al (2020). This work drew upon the semantic, syntactic, and typographic indicators in the original text, and has been expanded since this initial reporting. As anticipated, the SKOS standard limited the ability to encode many of the more complex logical connections Chambers' knowledge system presents. To explore in greater depth this complexity, the MRC is also producing a Web Ontology Language (OWL) encoding of the taxonomical tree and domain vocabulary to offer a separate basis of comparison. Both the SKOS and OWL versions of Chambers' system have been modeled from a naïve standpoint, where the modeler is not an expert in the area and has not had explicit recourse to a specialist, in this case one studying 18[th] century literature.

This paper explores the expressive and functional differences between the nearly complete SKOS and proposed OWL versions of Chambers' knowledge structure. As part of this goal, research has focused on the construction and application of rules in each system to produce a more computationally ready representation of Chambers' structure in SKOS, which is more thesaurus-like, and OWL, which represents additional ontological nuances. To explore the SKOS and OWL encoding processes in greater detail, this paper will analyze the modeling and syntactic aspects. To understand better the relationship between modeler and vocabulary this paper will evaluate Paul Grice's theory of conversational implicature as a possible theoretical framework for naïve modeling. The sections that follow first give an overview of Chambers knowledge structure; next explore constructs of terminology; explore modeling of the thesaurus and ontology; and finally, introduce Grice's theory of implicature.

**Overview of Chambers' *Cyclopaedia***

First published in 1728, Ephraim Chambers' *Cyclopaedia* attempted to gather the extant knowledge of the early $18^{th}$ century into a single volume. The work grew out of Chambers' desire to improve upon a previous encyclopedia produced by John Harris, *Lexicon Technicum*, which he worked on as an apprentice printer (Kennedy, 2013). The resultant work covers a wide range of subjects with entries similar to contemporary encyclopedias. To facilitate navigation of his work, Chambers offered a "Preface" that not only lays out his intentions for the overall project but also diagrams the various domains of knowledge he desires to represent. This knowledge map is divided into two separate but interconnected sections: first a *taxonomic tree* and second a *domain vocabulary*. The lowest nodes of the tree comprise the headwords for the vocabulary with a few exceptions (see Figure 1). The taxonomic tree bears similarity to Francis Bacon's "The Platforme of the Desygne" in *Of the Advancement and Proficience of Learning, or the Partitions of Sciences, IX Books* and was an inspiration for "Diderot's Tree of Knowledge." Chambers' iteration of the tree places "Knowledge" at the top level and sub-divides it into categories such as "Natural" and "Artificial" until reaching a level of granularity that results in a field of top concepts. He arranges and organizes the various aspects of the 47 top concepts he identifies to produce the domain vocabulary (Figure 2). Chambers offers a succinct, yet rich description of the goal of these two knowledge structures:

> it will be here necessary to carry on the Division of Knowledge a little further; and make a precise Partition of the Body thereof, in the more formal way of analysis: The rather, as an Analysis, by shewing the Origin and Derivation of the several Parts, and the Relation in which they stand to their common Stock and to each other; will assist in restoring 'em to their proper Places and connecting 'em together" (Chambers 1728, ii).

This process focuses on the breaking down of knowledge and reconstituting it into individual domains, listed in the "Preface." The features, objects, aspects, etc. of the domains are then enumerated and allow users to reference individual terms which appear in the main text of the *Cyclopaedia*.

Figure 1. Taxonomic tree from Chambers' "Preface."

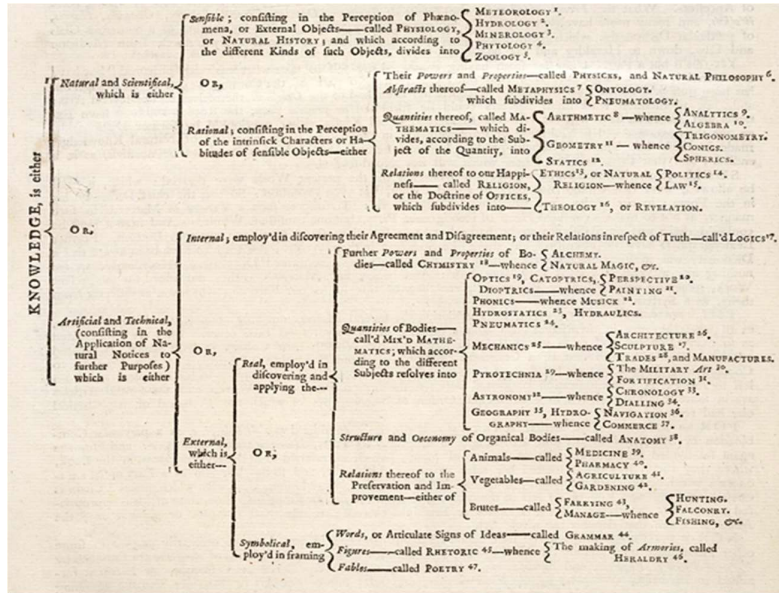

Figure 2. Example of domain vocabulary for "Minerology."

³ MINEROLOGY, or the History of EARTH; 1°, Its Parts, as *Mountain, Mine, Mofs, Bog, Grotto*; and their Phænomena, as *Earth-quake, Volcano, Conflagration*, &c. Its Strata, as *Clay, Bole, Sand*, &c. 2°, Foſſils or Minerals, as *Metals, Gold, Silver, Mercury*, &c. with Operations relating to 'em, as *Fuſion, Refining, Purifying, Parting, Eſſaying*, &c. *Litharge, Lavatory, Pinea*, &c. Salts, as *Nitre, Na-tron, Gemma, Allum, Armoniac, Borax*, &c. Sulphurs, as *Arſenic, Amber, Ambergreaſe, Coal, Bitumen, Naphtha, Petrol*, &c. Semi-metals, as *Antimony, Cinnabar, Marcaſite, Magnet, Biſmuth, Calamine, Cobalt*, &c. Stones, as *Marble, Porphyry, Slate, Asbeſtos*, &c. Gems, as *Diamond, Ruby, Emerald, Opal, Turcoiſe*, &c. *Emery, Lapis*, &c. whence *Ultramarine, Azure*, &c. Petrifactions, as *Cryſtal, Spar, Stalactites, Trochites, Cornu Ammonis*, and the like.

**Terminology**

The terms "subject matter expert" and "naïve" have variable and contested meanings. For the purposes of this paper, subject matter expertise relates to an individual's proficiency concerning the information under consideration and not that of their ability to model the subject. Similarly, the usage of the term "naïve" refers to the modeler's degree of understanding of how to express effectively the subject under consideration[1]. Both terms exist on a continuum with degrees of proficiency that range from the novice to the professional. The level of naivete regarding the modeling task could be influenced not only by one's own ability, but also by degree of access to a subject matter expert with which to consult. Other aspects that affect expertise are crossover skills obtained through coursework or employment experience which could influence a modeler[2].

**Modeling**

Modeling Chambers' vocabulary took two separate approaches that are currently at different stages of completion. The SKOS thesaurus is nearing completion while the OWL ontology is in a much earlier stage of development. Both transformations of the vocabulary largely derive from a similar base set of concepts, the domain vocabulary, but the ontology specifically engages the taxonomic tree structure which presents several distinct conceptual issues to overcome. In its attempts to engage both of Chambers' knowledge structures, the ontology enunciates both the class constructions and relationships between and across categories of knowledge. Work on the thesaurus continues that described in Greenberg et al. (2020).

**Thesaurus**

The thesaurus transformation project occurred in several distinct phases that each allowed for progressive refinement of Chambers' vocabulary. To structure the thesaurus, we followed the guidelines in ANSI/NISO Z39.19-2005 (R2010), "Guidelines for the Construction, Format, and Management of Monolingual Controlled Vocabularies." The first phase involved the basic sorting of headwords and associated terms into a spreadsheet, which were then compared with images from the "Preface." The most notable textual features of the vocabulary are typographic that evince shifts in font style between roman and italic, the use of dashes, the shorthand "&c." for et cetera, and the occasional use of numbered lists. These features each described aspects of the hierarchical organization within a conceptual framework, a fact that Chambers himself describes as, "EVERY Word is supposed to stand for some Part, or Point of Knowledge" (vi). With the exception of headwords and introductory equivalent terms, which have a distinct descriptive function, most of the words that appear in italic font refer to concepts that can be located within the *Cyclopaedia* while words or phrases in roman font serve a connective function, such as introducing a subcategory or aspect of the term under consideration as with the term "Law" which contains "publish'd in *Act*, *Statute*, *Charter*, *Rescript*, *Constitution*, *Decretal*, *Senatus-consultum*, *Pragmatic Sanction*, &c." (iv).

---

[1] This usage differs somewhat from Beghtol (2003) which describes naive classification systems as follows, "In contrast to information retrieval classification systems that support an environment in which searchers look for recorded knowledge, naïve knowledge discovery classifications support a scholarly environment in which new questions are expected to be asked of primary research materials" (p. 65). Chambers' system is already quite well formed, but the means of transforming it into a contemporary knowledge organizing system is somewhat different. The work on the OWL ontology might engage with this topic but is ultimately outside the scope of this paper.

[2] I drew upon work in classical languages, English grammar, and prior experiences in the editorial field while working on Chambers' domain vocabulary.

Moreover, in initial discussions, it was decided to treat such categories as facets of the headwords and represent them hierarchically. To terminate lists, Chambers often employs either a single period or the "&c." notation that generally denotes the shift to a new facet or else a possible separate grouping of concepts within a facet. Determining how to handle such shifts relies upon the associative relationships between terms before and after such breaks. Definitive shifts between ideas within a domain are often signaled by a long dash, space permitting, that is sometimes shortened at the end of a line. This latter notation is less common than other typographic conventions. As an example, in the Theology domain, Chambers uses the long dash to identify the shift between terms related to specific religions. These typographic elements allowed for quick identification of the structure within individual domain vocabularies.

And yet typography displays limits when the thesaurus model requires more meaningful or else more nuanced connections between concepts to be made, which the text constructs via a semantic framework. As seen with the earlier example from Law, the shift can be seen in font style, but the substance of the shift is conveyed through the meaning of published in which then controls the succeeding terms. Some of these terms are nested in a fashion to produce a specific order, such as in the term Geometry the description of measurements which are then connected to tools that make them. These semantic notations allow for thesaurus connections such as broader term (BT), narrower term (NT), and related term (RT) relationships to take shape. The internal semantic structure of domains tended to follow a specific pattern where Chambers introduced after the headword either an alternate name for the domain which was indicated by the word "or," or else he denoted the areas studied using the word "including." To capture as much information as possible, the alternate name or area under consideration was listed and the individual aspects, often denoted by small capital letters, were separately listed as either related or broader terms depending on whether they were standalone concepts or functioned as broader concepts within the overall structure. Facets within the individual domain areas developed from a range of syntactic and semantic cues. Internally Chambers used connective phrases not only to shift between topics but also to register instances, e.g., the word "as" is often utilized for such transitions.

Taken together, the structural features allow for Chambers to express series of highly idiosyncratic facets that often reflect the complexity of a domain and the absolute size of its terms. Moreover, while some specific facets seem to recur, e.g., "Operations," and have an implicit similarity they acquire contextual aspects from the domain in which they appear, a geometric operation bears little resemblance to medical one. Disambiguation of concepts, i.e., identical words which appeared under multiple terms, was the final operation performed on the vocabulary, where individual terms were cross referenced with their full definitions in the main text of the *Cyclopaedia*. Comprising approximately 250 concepts, this was the most semantically oriented aspect of the project that required a deeper engagement with Chambers' intended meaning. Terms that were used identically or whose semantic dissimilarity was slight were merged together, while others were differentiated by the domain in which they appear.

## Ontology

The facet structure and subcategorization process to produce the thesaurus underpins one element of the expected ontology; while engagement with the taxonomic tree forms the second primary element. Chambers' bifurcated knowledge structure offers up several possibilities and hazards. Working top down from the taxonomic tree, a variety of more traditional Aristotelian categories emerge that Frické (2012) characterizes as Aristotelian-Linnaean, displaying "JEPD (jointly exhaustive pairwise disjoint) property and it is possessed by Aristotelian classifications. [If JEPD holds for the leaves of a tree, it also holds for all the child nodes for any of the internal, non-leaf, parent nodes within the tree.]" (168). The tree structure encapsulates abstract categories that situate individual domain categories relative to each other within a universal structure which stems from "Knowledge," and the tree itself displays the JEPD property.

Two possible routes exist for modeling this structure. The first is to directly map Chambers' knowledge structure into an ontological one, e.g., where owl:thing is equivalent to chambers:knowledge and proceed down each branch. This interpretation of the taxonomic tree retains the greatest fidelity to Chambers' initial conception of knowledge and would integrate well with the lower-level domain ontology. However, there is the possibility of logical inconsistencies being introduced with identical terms at the lowest levels appearing in multiple branches of the tree, violating the jointly exclusive, pairwise disjoint structure. A second option for modeling the branches involves constructing axioms that describe the features and logical structure of the different levels of knowledge. This method allows for a much more descriptive ontology that can accommodate possible exceptions that occur at lower levels by producing a more polyhierchical framework. This approach also attempts to leverage the description logic features of the OWL standard possibly to enhance the use of the structure as a more general classification mechanism.

## Implicature

Implicature, generally, describes a theory that attempts to understand how participants in a conversation construct meaning from each other's statements, understanding what is implied based upon not only the statements but also context, tone, environment, etc. In "Logic and Conversation," Paul Grice describes the process as an attempt to understand problems expressed in natural language which cannot be equated to formal logical arguments (1989, 23-24). Given the relative lack of subject matter expertise of the primary modeler for this project, constructing a useful thesaurus and, as yet, framing a coherent ontology have presented unique challenges. One possible framework that helps to elucidate the relationship is Paul Grice's theory of implicature, which traditionally applies to participants navigating meaning in a conversational setting. Given the absence of a second conversant, i.e., Chambers himself, implicature might offer limitations for how extensively meaning can emerge outside of the ability to clarify his statements. Grice further describes the theory as follows, "Our talk exchanges do not normally consist of a succession of disconnected remarks and would not be rational if they did. They are characteristically, to some degree at least, cooperative efforts; and each participant recognizes in them, to some extent, a common purpose or set of purposes, or at least a mutually accepted direction" (Grice 1989, 26). Implicature requires the actors to participate in the exchange as a series of enunciations that each side attempts to navigate. In the case of a written text, one of the primary questions is whether the text can stand in for a speaker, in our case could the modeler and Chambers be engaged in a conversation, with each bringing their own assumptions to the information at hand? In fact, Grice's texts leave open the possibility for exchanges, in the form of "utterances," to be produced under a variety of circumstances that may even include non-verbal acts (1989, pp. 92, 118). If we extend this to include textual objects such as Chambers' vocabulary list, the goal of the modeler is to develop a viable knowledge structure that represents the intentions presented by the author.

One particular approach that Grice recommends is a "Modified Occam's Razor, *Senses are not to be multiplied beyond necessity*" (p. 47). This focus on engaging the language only insofar as what is given within the particular context lends itself to a highly limited interpretive view of language, which attempts to stem the proliferation of meaning beyond what one hears, or in this case reads. In the Chambers' vocabulary the concepts themselves are fixed elements and mostly lack innate meaning[3] within the context of the domains presented beyond their capacity to offer a link to information within the greater *Cyclopaedia*. The words that provide meaning within the domains are notably the connective phrases noted in the earlier sections on modeling. Modeling Chambers' text required situating these phrases in relation to both the overarching domain and in relation to other phrase structures within the domain, but it is a functional understanding that is gained to determine how concept groups fit together rather than constructing an interpretive apparatus from without the text. In rare instances, connections were interpolated when a noticeable shift had occurred, e.g., in the "Heathen" grouping from Theology that shows distinct a distinct shift from "gods" to a separate aspect that was considered different enough to add interpolate a heading for rites (see Figure 3). In this situation the shift in senses of the concepts either warranted a synthetic heading or else placing the terms in direct relation as narrower terms to Heathen. Because there was precedence for a facet called rites under the Christianity group, Chambers' intention, while unstated, seems clear. However, locating such shifts relies upon the modeler noting them, which often necessitates attention to word forms, e.g., shifts from plural to singular.

Figure 3. Note that "Gods" was considered to conclude at "Genius" and a transition for Rites was interpolated as beginning at "Apotheosis."

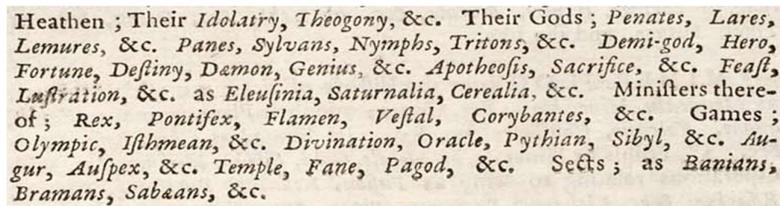

---

[3] The typographic and syntactic forms of the concepts can denote subtle shifts between or within facets. Also, obvious changes in subject matter within a domain should prompt some reflection by the modeler; however, there are often other markers which denote a shift, such as the terminal &c. or simply a period.

**Discussion and Analysis**

Modeling Chambers' vocabulary has presented a variety of issues that have primarily focused on the construction of the thesaurus. At a basic level the text offers a highly structured set of terms which forms the basis of a SKOS thesaurus that can be used for computational and automatic indexing purposes. The process of transforming the vocabulary focused heavily on how modeler and text interact.

One aspect which requires some further examination is the use of implicature for constructs outside of the more basic thesaurus construction. Implicature seems to envision a controlled set of circumstances and conventions between two parties to make sense of the others' intended meaning. Thesauri express a rather narrow set of relationships among concepts, which can be understood by an author's conventions, such as syntax and typography. But implicature might be tied heavily to complexity of the resulting knowledge structure. Given the preparatory work done on the ontology of Chambers' whole knowledge structure, implicature helps to form the basic classes which map closely to the facets. However, to model the taxonomic tree and the connections to the domains a more interpretive approach may be necessary which requires more hermeneutic features. The reason for this is twofold. First the level of abstraction necessary for metamodeling ideas means intuiting structures which may exceed the author's intention, e.g., features or trends noticed by the modeler but not explicitly or implicitly suggested by the text. Second, the structure of an ontology may invite the use of polyhierarchies or even utilize aspects of external ontologies to classify concepts whose logic exists without the knowledge organization at hand. Ultimately the thesaurus allows for a greater degree of fidelity to the text while an ontology offers a more expressive conceptual view of knowledge structure because it engages the taxonomic tree.

In conjunction with the role of implicature the twin concepts of naivete and subject matter expertise become somewhat problematic. Subject matter expertise displays a high degree of variance for modeling a vocabulary such as Chambers. A greater degree of proficiency would probably locate connections or understand Chambers' intentions better than a naïve modeler. But what this paper shows is that modeling occurs along a wide range of skills which are attuned to different aspects of the vocabulary. Attention to structural features and syntactic elements of the text, in this case, have an outsized influence on the capacity to locate facets and features of the vocabulary. Moreover, naivete in regard to the subject at hand, here 18th century literature, might be offset by adjacent skills in language, science, classical literature, or associated literature studies. In the case of Chambers' vocabulary, subject matter expertise in information science was readily available, but access to an 18th century literature scholar was limited. Regardless, a tool such as implicature might enhance outcomes of vocabulary modeling for more basic structures if only to provide individuals working with the text a sense of connection with material that might be alien.

**Conclusion**

The research presented here explored the Chambers' ontology and to make it machine-readable. Overall, the work shows the complexity for a naïve modeler to implement an effective machine-readable thesaurus or ontology. Further steps in this project include the completion of the thesaurus and the ontology. For the former, there are minor editorial aspects which need to be attended to. For the ontology, the specifications and structure are in process and will eventually be constructed using the Protege application. A last aspect of this project involves the assignment of permanent digital identifiers to the concepts to ensure longevity of the overall project. Among the possible outgrowths of this project might be a set of best practices for naïve modeling for situations similar to those presented by the Chambers' vocabulary.